\documentclass[12pt]{article} 
\usepackage{epsf}
\usepackage{color}

\begin{document}
\begin{center}
{\bf \Large Dissociation of a boosted quarkonium in quark-gluon plasma}
\end{center}
\vspace*{1cm}
\begin{center}
{\bf Sidi Cherkawi Benzahra\footnote[1]{sidi.benzahra@ndsu.nodak.edu}}
\end{center}
\begin{center}
{\large North Dakota State University}
\end{center}
\begin{center}
{\large Fargo, North Dakota 58105}
\end{center}
\vspace*{1cm}
\begin{abstract}
I consider the dissociation of a boosted quarkonium in a quark-gluon plasma.  
This dissociation is due to absorption of a thermal gluon.  I discuss the 
dissociation in terms of the velocity of the quarkonium and the temperature of
the quark-gluon plasma.  I compare this dissociation rate to the one calculated
without including the velocity of the quarkonium.

\end{abstract} 
\vspace*{1cm}
  
\hspace*{.5cm}In a quark-gluon plasma, the quark or the anti-quark of the quarkonium can be struck 
by a high energy gluon and  
the quarkonium can dissociate into other elements [1].  The medium, the quark-gluon plasma, is 
full of gluons that can cause this dissociation, and this can happen by 
exciting a color-singlet ${\mid q\bar{q} \rangle}^{(1)}$ into a
color-octet continuum state${\mid q\bar{q} \rangle}^{(8)}$ [2]. The quark absorbs 
energy from the gluonic field of the quark-gluon plasma, but there is a 
threshold energy that the gluon has to have in order to dissociate the quarkonium
in the quark-gluon plasma. For example, in free space, there is a threshold energy of about 
850 MeV [3] for the dissociation of the upsilon meson into two highly-energetic 
bottom quarks.  I choose the upsilon meson, $\Upsilon$, for simplicity, since it is heavy and its 
binding energy is large compared to other mesons such as the pion or the $J/\psi$ which 
dissociate fast and at relatively low temperature [4].  Moreover, in an experiment which 
produces a very high temperature quark-gluon plasma, the $J/\psi$ will break down at a faster 
rate compared to the $\Upsilon$ and only a few of the $J/\psi$'s will survive.  This makes it hard for 
the experimentalist to detect the $J/\psi$ in order to confirm that a quark-gluon plasma had 
materialized. On the other hand, the $\Upsilon$ will survive the high temperature of the  
quark-gluon plasma and therefore it will be easy to detect. Moreover, the upsilon meson is small in size 
compared to the $J/\psi$ so it needs a higher density plasma for it to dissociate.  In this 
kind of dissociation, we state that the $\Upsilon$ is no longer bound in the singlet state.  In quark-gluon 
plasma the threshold energy will become less than the one in free space. Only gluons 
with energy exceeding the threshold energy in quark-gluon plasma can dissociate the quarkonium.  
So the relevant energy density is not just the average energy density, but the 
energy of the gluons which have an energy higher than the threshold energy.  
In deconfined matter, such as quark-gluon plasma, we expect gluons in a medium 
of 200 MeV temperature to have an average energy of 600 MeV [5], which can dissociate 
the quarkonium.\\

\hspace*{.5cm}The effect of screening can also cause the meson to dissociate. 
A screen of quarks and gluons builds up between the quark and the anti-quark 
of the quarkonium and the two quarks cannot feel each other's attraction and the 
quarkonium dissociates.  This kind of screening is called Debye screening [6] 
and I discussed it in a previous work [7]. In regards to screening, once we take 
the motion of the quarkonium into consideration, the dielectric properties of the 
medium [8], which is the quark-gluon plasma, gets introduced and the 
physics becomes somewhat involved.\\

\hspace*{.5cm} When the heavy quark and anti-quark of the quarkonium are close
to each other, asymptotic freedom comes into play, and the binding energy
can be derived the same way we derive it for the hydrogen atom. To an 
approximation, there is a parallelism between the two physics [2].\\  

The wavelength of the gluon that dissociates this state fits the radius of
the singlet state of $\Upsilon$ to a good approximation. 
The S matrix in this case is:
\begin{equation}
{S_{fi}}=-i\int_{-\infty}^ {\infty} dt \langle {\rm octet} \mid grE^{a}\cos\theta
\mid 
 {\rm singlet} \rangle 
\end{equation}
where $E^{a}$ is the color electric field and $g$ is the color charge.  It is worked out 
in [2,7] that the dissociation rate of the upsilon meson becomes:
\begin{equation}
\Gamma_{dis}\approx {2\over3}\pi^{3} {\alpha_{s}} a^2 n \, , 
\end{equation} 
where $a$ is the radius of the singlet state of the quarkonium, and $n$ is the number density of gluons
in the quark-gluon plasma.  Since we are considering thermalization, this dissociation rate can be calculated in terms of the temperature of the
quark-gluon plasma. But I will only include the gluons with energy exceeding 
the threshold energy of dissociation, $\omega_{\rm min}$.  For a medium of gluons
\begin{equation}
n=N/V={1\over{2\pi^2}} \int_{\omega_{\rm min}}^{\infty} {d\omega} {{\omega^2}\over 
{{\rm exp}{(\omega/T})}
 -1} \, ,
\end{equation}

which gives us
\begin{equation}
\tau_{dis} \approx {{4\alpha_{s}}\over{3\pi}}{{{m_{Q}^{2}}}\over{
\sum_{k=1}^{\infty} 
\left[ \left ({{T}\over{k}} \right )\omega^2_{\rm min}
+2{\left ({{T}\over{k}}\right )^2}
\omega_{\rm min}+ 2 \left ({{T}\over{k}}\right )^3 \right ]
e^{-k\omega_{\rm min}/T}}} \, ,
\label{lifetimewithoutv}
\end{equation}

where $\tau_{dis}$ is the dissociation time of the quarkonium.  The minimum temperature required 
to achieve deconfinement is generally understood to be about 150 to 200 MeV.  RHIC, the Relativistic Heavy Ion 
Collider, is the first collider designed to specifically create this plasma.  It
may reach a temperature of 500 MeV.  CERN is trying to reach a temperature of 1 GeV by 
colliding heavy nuclei in the Large Hadron Collider.  Inserting a temperature of 500 MeV and a 
threshold energy $\omega_{\rm {min}}$ of 10 MeV into eq.(\ref{lifetimewithoutv}), I get a dissociation time of 
23 fm/c, which is somewhat comparable to the lifetime of the quark-gluon plasma--a typical lifetime of a
quark-gluon plasma is about 10-20 fm/c [9,10].  If I use the CERN 1 GeV 
temperature, I get $\tau_{\rm dis}\approx 3\; \mbox{fm/c}$. We should remember that I used 
the threshold energy of 10 MeV as the binding energy of the upsilon quarkonium. 
This is due to screening, and to the fact that the medium is hot.\\

In the case of a moving quarkonium, we have to consider the
velocity when calculating for the dissociation time.  And since the quarkonium  
is moving in the frame of the gluons, the gluons will be relatively moving in the frame of the 
quarkonium. So in this case, the momentum, $\omega$, of the gluon will change into 
${\omega^{\prime}} = \gamma \omega (1-v\rm cos\theta),$ where $\theta$ is the angle between the velocity vector of the meson and the momentum vector of the gluon, and $v$ is the 
relative velocity of the quarkonium in the quark-gluon plasma. The new number density 
of gluons in quark-gluon plasma, in this case, is:

\begin{equation}
{n^{\prime} }={1\over{2\pi^2}} \int_{\omega_{min}}^{\infty}{d\omega} \int_{1}^{-1} {d(\rm cos \theta)} {{\omega^2}\over 
{\rm exp ({{{\gamma \omega (1- \beta \rm \cos\theta)}}\over {T}} }
 ) -1} \, ,
\end{equation}

Integrating the number density over $\omega$ and $\rm \cos \theta$, we get a dissociation time of the form
\begin{equation}
\tau_{dis} \approx {{4\alpha_{s}}\over{3\pi}}{{{m_{Q}^{2}}}\over{
\sum_{k=1}^{\infty}}
\left \{f(v, T) \left ({{1} \over {\beta}}\right ) {\rm sinh}({{k\gamma \beta \omega} \over {T}}) + g(v, T) 
{\rm cosh}({{k\gamma \beta \omega} \over {T}}) \right \}{{e^{-{k \gamma \omega_{\rm min} \over{T}}}}} }\, .
\label{lifetimewithv}
\end{equation}

where 

\begin{equation}
 f(v, T)= \left\{ \left ({{T}\over{k}} \right )^{2} \omega_{\rm min}
+ \gamma \left (1+{\beta}^{2} \right ){\left ({{T}\over{k}}\right )^3} \right\} \nonumber
\end{equation}

and 

\begin{equation}
g(v, T)=\left\{ \left ({{T}\over{k}} \right )^{2}\omega_{\rm min}
+ 2 \gamma {\left ({{T}\over{k}}\right )^3} \right\} \nonumber
\end{equation}

The latter equation of the dissociation time indicates that the quarkonium breaks down faster 
when its velocity increases in the quark-gluon plasma--we have to be careful here because this 
might not happen in the case of screening.  Although the velocity of the quarkonium 
is small, its dissociation time with respect to its velocity should not be ignored.  For 
example, at a quark-gluon plasma temperature of 1 GeV, a non-moving quarkonium dissociates 
in a time duration of 3 fm/c, but if its velocity reaches 0.5 c, for instance, its dissociation time will 
decrease to 2.64 fm/c. If a quarkonium can reach a velocity of 0.9 c in 1 GeV quark-gluon 
plasma, which is unlikely to happen, because the transversal momentum, $P_{T}\sim v\gamma 
m_{\rm{meson}}$ can only be in the range of 5 GeV [11] due to beam constrains, the quarkonium 
dissociates in a time duration of 1.48 fm/c.  It is easy to check that in the limit $v \rightarrow 0,$ the 
dissociation time of 
Eq.(\ref{lifetimewithv}) reduces to that of Eq.(\ref{lifetimewithoutv}).\\

In conclusion, the quarkonium binding energy weakens with the increase of the 
temperature of the quark-gluon plasma and with the increase of the velocity of the quarkonium.  
But the velocity effect is smaller compared to that of the temperature. Thermalized 
gluons can play a major role in dissociating quarkonia, but since hard gluons can also be 
produced in quark-gluon plasma [12], dissociation of quarkonia due to collision with hard 
gluons needs to be investigated.  This is a tricky problem though because we have to take 
into consideration the expansion of the plasma, and we cannot make use of the temperature 
because hard gluons escape the quark-gluon plasma before thermalization. 
\begin{center}
{\bf \large Acknowledgment}
\end{center}
I am indepted to Joseph Kapusta and Berndt Mueller for their help and advice on this paper. I am also 
indepted to Benjamin Bayman for his general help.  I also want to 
thank Steve Estvold, Scott Tollefson, Claudio Verdozzi, and Anthony Wald for their help on this paper.\\

\begin{center}
{\bf \large References}
\end{center}
[1] K. Tsushima, A. Sibirtsev, K. Saito, A. W. Thomas, and D. H. Lu, Nucl. Phys. (2000) 279-284. \newline
[2] B. Muller, preprint nuc-th/9806023, v2.\newline
[3] Review of Particle Physics, Volume 15, Number 1-4, 2000. \newline
[4] T. Matsui and H. Satz, Phys. Lett. B $\bf 178$ (1986) 416. \newline
[5] D. Kharzeev and H. Satz, Nucl. Phys. A $\bf 590$ (1995) 515c-518c .\newline
[6] H. Satz, Nucl. Phys. A $\bf 418$ (1984) 447c. \newline
[7] S. C. Benzahra, Phys. Rev. C $\bf 61$ 064906 (2000).\newline
[8] M.C. Chu and T. Matsui, Phys. Rev. D $\bf 39$, 1892 (1989).\newline
[9] S. Y. Wang and D. Boyanovsky, Phys. Rev. D $\bf 63$ (2001) 051702 . \newline
[10] J. W. Harris and B. Muller, Annu. Rev. Nucl. Part. Sci. $\bf 46$, 71 (1996). \newline
[11] D. Kharzeev and H. Satz, Physics Letters B 334, 155 (1994). \newline
[12] D. H. Rischke and M. Gyulassy, Nucl. Phys. A $\bf 597$ (1996) 701-726 .\newline
 
\end{document}